\documentclass[useAMS]{mn2e}
\usepackage{graphicx}

\begin{document}

\title[Asteroseismology of $\nu$~Eridani: Interpretation and applications]
{Asteroseismology of the $\beta$ Cephei star $\nu$ Eridani:
  Interpretation and applications of the oscillation spectrum}
\author[A. A. Pamyatnykh et al.]
  {A. A. Pamyatnykh$^{1,2,3},$\thanks{E-mail: alosza@camk.edu.pl}
G. Handler$^{3}$ and W. A. Dziembowski$^{1,4}$
\and \\
$^1$ Copernicus Astronomical Center, Bartycka 18, 00-716 Warsaw, Poland\\
$^{2}$ Institute of Astronomy, Russian Academy of Sciences,
Pyatnitskaya Str. 48, 109017 Moscow, Russia\\
$^{3}$ Institut f\"ur Astronomie, Universit\"at Wien,
T\"urkenschanztrasse 17, A-1180 Wien, Austria\\
$^{4}$ Warsaw University Observatory, Al. Ujazdowskie 4, 00-478 Warsaw,
Poland}

\date{Accepted 000.
  Received 2004 February 000;
  in original form 2004 January 5}
\maketitle
\begin{abstract}

The oscillation spectrum of $\nu$ Eri is the richest known for any
variable of the $\beta$ Cephei type. We interpret the spectrum in terms of
normal mode excitation and construct seismic models of the star. The
frequency data combined with data on mean colours sets the upper limit on
the extent of overshooting from the convective core. We use data on
rotational splitting of two dipole ($\ell=1$) modes (g$_1$ and p$_1$) to
infer properties of the internal rotation rate. Adopting a plausible
hypothesis of nearly uniform rotation in the envelope and increasing
rotation rate in the $\mu$-gradient zone, we find that the mean rotation
rate in this zone is about three times faster than in the envelope. In our
standard model only the modes in the middle part of the oscillation
spectrum are unstable. To account for excitation of a possible high-order
g-mode at $\nu=0.43\mbox{ cd}^{-1}$ as well as p-modes at $\nu > 6\mbox {
cd}^{-1}$ we have to invoke an overabundance of Fe in the driving zone.

\end{abstract}


{\section{Introduction}}

The rich oscillation spectrum of $\nu$ Eri obtained in a recent multisite
campaign (Handler et al. 2004, Aerts et al. 2004) holds the best prospects
for seismic sounding of the interior of a B-star but also presents a
considerable challenge to stellar pulsation theory. The sounding is
facilitated by the fact that for several excited modes the spherical
harmonic indices were determined (De Ridder et al. 2004). The challenge is
to explain mode excitation in an unexpectedly broad frequency range.


The oscillation spectrum of $\nu$ Eri is shown in Fig.\,1. Before the
campaign only the $\ell=0$ mode and the $(\ell=1,\mbox{ g}_1)$ triplet
were known. However, for the triplet, the identification of the spherical
harmonic was uncertain. Now, thanks to the analysis of accurate
multicolour photometry and high-resolution spectroscopy, the uncertainty
was eliminated. Reliable information about $\ell$ is a crucial input for
constructing seismic models of a star, that is, the models constrained by
the frequency data.

\begin{figure}
\begin{center}
\includegraphics[width=88mm,viewport=05 30 555 381]{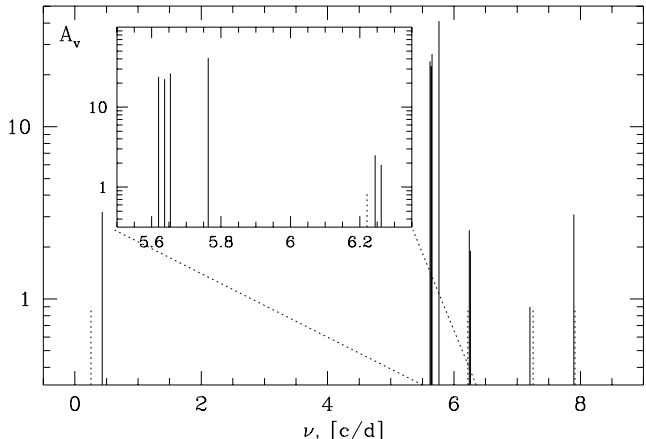}
\end{center}
\caption{Oscillation spectrum of $\nu$ Eri. The peaks represented with the
broken vertical lines are regarded uncertain. 
The four modes known before the campaign are between $\nu=5.6\mbox{ and
5.8 cd}^{-1}$. De Ridder et al. (2004) identified them as an $\ell=1$
triplet
and an $\ell=0$ singlet. The newly discovered triplet at
$\nu\approx6.25\mbox{
cd}^{-1}$ is identified as $\ell=1$ (De Ridder et al. 2004). In this paper
we
will show that the radial mode is the fundamental and the two dipole modes
are
${\rm g}_1$ and ${\rm p}_1$, respectively.
Only these mode frequencies are used
in our seismic sounding.}
\end{figure}

In this work, we attempt to make the best use of the frequency data to
address unsolved problems in stellar evolution theory concerning element
mixing in convectively stable layers and angular momentum evolution. The
problems are related because rotation induces fluid flows that may cause
mixing. It is important to disentangle effects of overshooting from the
convective core and effects of rotationally induced element mixing. The
first effect depends only on stellar mass. The second one depends on
randomly distributed stellar angular momenta. In this context, $\nu$
Eridani, which is a very slowly rotating B star, yields a useful extreme
case. However, it would be important to try to obtain similar quality data
for more rapidly rotating objects.

We are very curious how the angular velocity of rotation behaves in
stellar interiors. Let us remind that, while the results of helioseismic
sounding of radial structure essentially confirmed the standard solar
model, the results for the internal rotation rate came as a surprise. The
outstanding question in the case of massive stars is whether a substantial
radial gradient of the angular rotation rate may be present in their
interiors. The question is related to another very interesting and
important problem which is the origin of magnetic fields in B type stars.

A seismic model of a pulsating star should not only account for the
measured frequencies but also for the instability of the detected modes.
The latter requirement sets a different kind of constraint than the former
as it concerns only properties of the outer layers, where most of
contribution to mode driving and damping arises. Since the frequency
spectrum of $\nu$ Eri seems unusually broad for an object of its type,
constraints in this case are particularly interesting.

The structure of this paper is as follows. In Sect. 2 we construct models
of the internal structure for $\nu$ Eri and calculate its oscillation
frequencies. We find models for which the oscillation frequencies of three
$m=0$ modes -- $(\ell=0\mbox{, p}_1)$, $(\ell=1\mbox{, g}_1)$ and
$(\ell=1\mbox{, p}_1)$ -- reproduce the observations as well as have
effective temperatures and luminosities within the observational error
box. In Sect. 3 we use these models and data on the triplet structure to
infer the internal rotation of the star.

The problem of how to make the observed modes pulsationally unstable is
discussed in Sect. 4. The tasks of constructing the model of $\nu$ Eri
whose mode frequencies match the data and of the identification of the
driving effect are not quite independent. The driving effect in B stars
strongly depends on the iron content and its distribution in outer layers.
In Sect. 5, we study how modification of the iron distribution affects the
frequencies. We also propose plausible identifications of all modes
detected in the $\nu$ Eri oscillation spectrum.

\vspace{4mm}

\section{Model of the stellar interior}

For the purpose of constructing models of $\nu$ Eri we ignored all effects
of rotation which, as we have checked, was fully justified. Accordingly,
we make use here only of the $m=0$ mode frequencies despite, in general,
rotation causes a shift of the centroid frequency. We consider models of
massive main sequence stars in the phase of radius expansion.  A model is
characterised by its mass, $M$, effective temperature, $T_{\rm eff}$,
initial hydrogen, $X$, and heavy element, $Z$, abundance, as well as the
overshooting parameter, $\alpha_{\rm ov}$. Here we assume the standard
solar mixture of heavy elements (departures will be considered in section
5) and fix $X$ at 0.7. For opacity and the equation of state we use the
OPAL tabular data (Iglesias \& Rogers 1996, Rogers et al. 1996). The
nuclear reaction rates are the same as used by Bahcall and Pinsonneault
(1995).

\subsection{$\nu$ Eridani's position in the theoretical H-R diagram}

The position of the star in the $\log T_{\rm eff} - \log L$ diagram is not
very accurately known. Our temperature determination is based on
Str\"omgren photometry and on the tabular data by Kurucz (1998). For $L$
we use the Hipparcos parallax, Lutz-Kelker correction to the absolute
magnitude (Lutz \& Kelker 1973), and the bolometric correction from
Kurucz's data.  In this way we obtained the following numbers $$\log
T_{\rm eff}=4.346\pm0.011\quad\mbox{ and }
\quad\log L=3.94\pm0.15.$$ For
$\log T_{\rm eff}$ we include measurement errors and uncertainty of
calibration. For $\log L$ we take into account errors of the parallax and
uncertainty of the bolometric correction. Our numbers are similar to those
quoted by Dziembowski \& Jerzykiewicz (2003) and will be updated by De
Ridder et al. (2004).

\begin{table*}
\begin{center}
\caption {Parameters of fitted models. The symbols denote : $X_{\rm c}$ --
hydrogen abundance in the core, $M_{\rm mix}$ -- mass of the mixed core,
which is the same as mass of the convective core at $\alpha_{\rm ov}=0$,
$r_{\rm c}$ -- radius of the convective core, $r_{\rm c0}$ -- radius of the
convective core at ZAMS, which corresponds to the top of the $\mu$-gradient
zone.}
\begin{tabular}{|cccccccccccc|}\hline
 $M/M_\odot$ & $\alpha_{\rm ov}$ & $Z$ & age[My] & $\log T_{\rm eff}$ &
 $\log L$ & $R/R_\odot$ & $\log g$ &
 $X_{\rm c}$ & $M_{\rm mix}/M$ & $r_c/R$ & $r_{c0}/R$ \\
\hline 9.858 & 0.0 & 0.015 & 15.7 & 4.3553 & 3.969 &
 6.279 & 3.836 & 0.2414 & 0.189 & 0.117 & 0.186 \\
\hline 9.179 & 0.1 & 0.015 & 18.9 & 4.3399 & 3.891 &
 6.165 & 3.821 & 0.2555 & 0.211 & 0.121 & 0.185 \\
\hline
\end{tabular}\\
\end{center}
\end{table*}

\subsection{Identification of the radial overtone of the observed modes}

Before starting seismic modelling of $\nu$ Eri, we must complete the mode
identification. The work by De Ridder et al. (2004) provided unambiguous
$\ell$ identifications for seven modes, and $m$ determinations by
inference, but the radial overtone of the modes has not been determined.
Fortunately, one of the detected pulsation modes is radial, which eases
this task considerably.

Consequently, we follow Dziembowski \& Jerzykiewicz (1996) and only
consider two one-dimensional families of models. The position of $\nu$ Eri
in the HR diagram is only consistent with the radial mode at $5.7634
\mbox{ cd}^{-1}$ being the fundamental or the first overtone (as indicated
later in Fig.\,3). For these two possibilities, we examine the mass
dependence of theoretical $\ell=1$ mode frequencies of standard models and
match it to the observed $\ell=1$ mode frequencies (Fig.\,2).

\begin{figure}
\begin{center}
\includegraphics[width=88mm,viewport=-5 20 560 545]{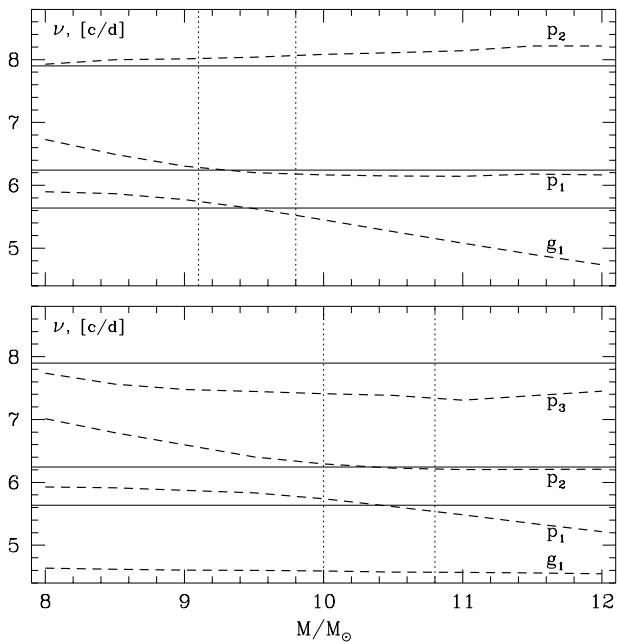}
\end{center}
\caption{Identification of the radial overtone of the observed modes of
$\nu$ Eri. We assume the radial mode to be the fundamental in the upper
panel and to be the first overtone in the lower panel. The mass dependence
of theoretical $\ell=1$ mode frequencies (dashed lines) is compared to the
observed $\ell=1$ mode frequencies (horizontal lines). We find two ranges
in mass (denoted by the vertical dotted lines) where the two $\ell=1$
modes of lowest frequency match the observed frequencies well. However,
the $\ell=1$ mode near $7.9 \mbox{ cd}^{-1}$ can only be reproduced in the
case of the radial mode being the fundamental.}
\end{figure}

Because of the mixed-type character of some of the modes we can only match
the observed frequencies of the two lowest-order $\ell=1$ modes within
certain mass ranges. However, under the assumption of the radial mode
being the first overtone, no acceptable fit can be found for the $\ell=1$
mode of highest frequency. Hence we can eliminate this possible
identification (which would also be less consistent with the star's
position in the HR diagram) and identify the radial pulsation mode of
$\nu$ Eri as the fundamental. Consequently, the $\ell=1$ modes are the
first g-mode (g$_1$) and the first two p-modes (p$_1$ and p$_2$). We note
that the theoretical fit of the standard model to the observed p$_2$ mode
is also not very satisfactory at first glance; we will return to this
problem and its possible solution later.

\subsection{Fitting frequencies of the $\ell=0$ and two $\ell=1, m=0$ modes}

We calculated frequencies of these three modes in evolutionary sequences
of stellar models characterised by the parameters $M$, $Z$, $\alpha_{\rm
ov}$. The first two of them and $T_{\rm eff}$ were regarded as adjustable
parameters to fit the measured frequencies. The adopted precision of the
fit was $10^{-3}$. The precision of the data is much higher but matching
to such a precision is neither interesting nor reasonable in view of the
uncertainties of the model calculations, such as the limited accuracy of
the opacity data. The upper limit of the $\alpha_{\rm ov}$ parameter was
derived from the bounds on the effective temperature.

We obtained models with $\alpha_{\rm ov}=0.0$ and $\alpha_{\rm ov}=0.1$
which satisfy all the observational constraints. The inferred value of the
metallicity parameter is $Z=0.0150\pm0.0001$, the same at $\alpha_{\rm
ov}=0.0$ and 0.1. Parameters of the two selected models are given in Table
1.

Both models are well within the error box in the theoretical H-R diagram
shown in Fig.\,3. Increasing $\alpha_{\rm ov}$ above 0.12 would place the
seismic model below the lower bound of $T_{\rm eff}$. We, thus, conclude
that the data on $\nu$ Eri are consistent with negligible overshooting
distance and set an upper bound on it at $\alpha_{\rm ov}=0.12$. Further
constraining the extent of overshooting may be possible when the effective
temperature determination is improved. Our constraint is in agreement with
$\alpha_{\rm ov}=0.10\pm0.05$ derived by Aerts et al. (2003) from
frequency analysis for the $\beta$ Cephei star HD 129929, which is also
slowly rotating.

\begin{figure}
\begin{center}
\includegraphics[width=88mm,viewport=-5 15 545 310]{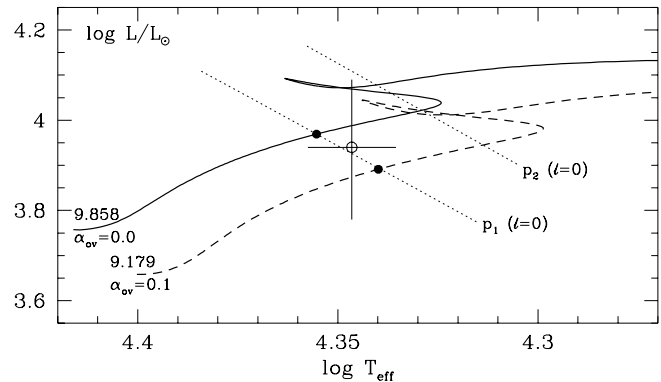}
\end{center}
\caption{The evolutionary tracks for the models fitting
frequencies of the $(\ell=0\mbox{, p}_1)$, $(\ell=1\mbox{, g}_1)$
and $(\ell=1\mbox{, p}_1)$ modes in $\nu$ Eri. Models were
calculated with chemical composition parameters $Z=0.015$
(adjusted) and $X=0.7$ (fixed). The values of $M$ and $\alpha_{\rm
ov}$ are indicated. The values of $T_{\rm eff}$ and $\log L$
inferred from photometry and the Hipparcos parallax are shown with
the error bars. Two dotted lines connect models
with 5.7633 c/d as the radial fundamental (p$_1$) and
first overtone (p$_2$) mode.}
\end{figure}
\smallskip

The frequency distance between the $(\ell=1\mbox{, g}_1)$ and
$(\ell=0\mbox{, p}_1)$ modes is indeed a very sensitive probe of
overshooting. This is illustrated in Fig.\,4, where we show the evolution
of the three mode frequencies as a function of $T_{\rm eff}$. This is the
main application of the $(\ell=1\mbox{, g}_1)$ mode. However, without the
$(\ell=1\mbox{, p}_1)$ mode frequency the assessment of $\alpha_{\rm ov}$
would not be possible unless we have an accurate observationally
determined $Z$ value which we do not.

\begin{figure}
\begin{center}
\includegraphics[width=88mm,viewport=-5 10 560 550]{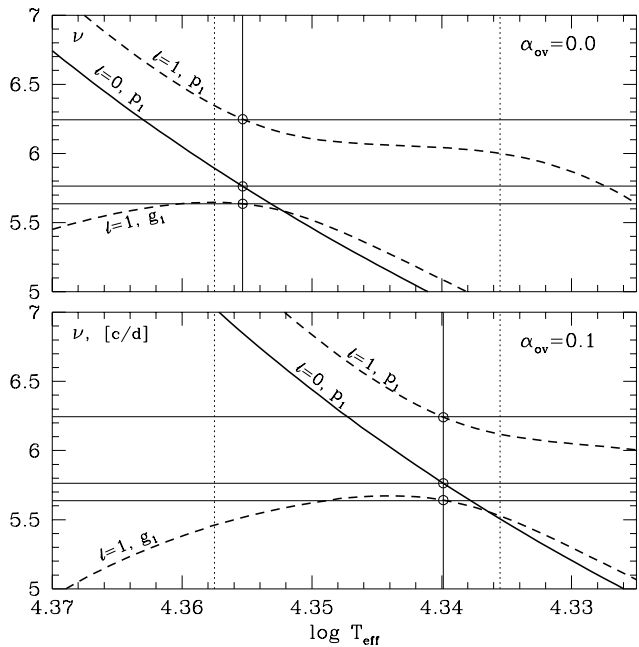}
\end{center}
\caption{The evolution of the calculated frequencies of the selected modes
in the model sequences shown in Fig.\,2. Two dotted vertical lines mark
the allowed $T_{\rm eff}$ range. Three horizontal lines correspond to the
measured frequencies. Solid vertical lines correspond to $T_{\rm eff}$ for
the fitted models. The fittings are marked by open circles. Note that the
g$_1$ mode frequency comes close to the radial mode frequency in a very
narrow range of $T_{\rm eff}$ and the location of this range is very
sensitive to $\alpha_{\rm ov}$. Crossing of the two frequencies takes
place at the point of the minimum frequency distance between the two
$\ell=1$ modes (avoided crossing). Before the avoided crossing, the
$(\ell=1\mbox{, p}_1)$ mode is nearly a pure acoustic mode with its
frequency evolution line being nearly parallel to the corresponding line
for the radial mode. The ratio of the two frequencies depends on $Z$.}
\end{figure}

%
\subsection{Probing properties of low order $\ell=1$ modes}

It is remarkable how much the three mode frequencies tell us about the
star's internal structure. Identification of a radial mode is important
because its frequency yields an accurate constraint on the model
parameters; however, much more interesting is the information contained in
the $\ell=1$ modes.

The probing property of a mode depends on the distribution of its
oscillation energy, $E$, in the stellar interior. The plots in the upper
panel of Fig.\,5 show $\cal E$, the energy derivative with respect to the
fractional radius, for the four consecutive $\ell=1$ modes. In spite of
rather close frequencies, modes g$_1$ and p$_1$ have a grossly different
distribution of energy. This explains the differences in their evolution
seen in Fig.\,4 as well as the difference in the probing properties. One
should also notice the peculiar ${\cal E}(r)$ dependence for the g$_1$ and
p$_1$ modes in the deep interior. A maximum (absolute in the former and
local in the latter case) is reached at the boundary of the convective
core where the Brunt-V\"ais\"al\"a frequency has a derivative
discontinuity. The g$_2$ and all higher order g-modes have the maxima
within the propagation zone. For our models it is the g$_1$ mode whose
frequency is most sensitive to overshooting. In more evolved models this
property will shift to p$_1$. Dziembowski \& Pamyatnykh (1991) observed
this property of certain low order g-modes, called by them g$_c$, in
$\delta$ Scuti star models, and emphasized the importance of detecting
such a mode in the context of the overshooting problem. $\nu$ Eri is the
first object in which a g$_c$ mode was definitely detected. More than
that, its rotational splitting has been measured.

\begin{figure}
\begin{center}
\includegraphics[width=88mm,viewport=-5 20 560 550]{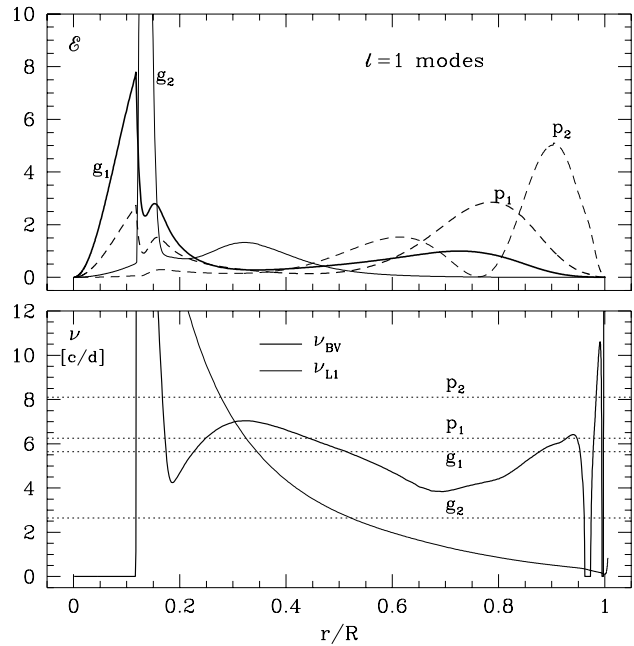}
\end{center}
\caption{ {\it Upper panel:} Oscillation energy distribution, $\cal E$, for
low order $\ell=1$ modes in the $\nu$ Eri model calculated with $\alpha_{\rm
ov}=0$. The plots are very similar in the models with $\alpha_{\rm ov}=0.1$.
{\it Lower panel:} The Brunt-V\"ais\"al\"a frequency, $\nu_{\rm BV}$, and
the Lamb frequency at $\ell=1$, $\nu_{\rm L1}$, in the same model.}
\end{figure}

$\nu$ Eridani is not exceptional amongst $\beta$ Cep stars, which are $9 -
12 M_\odot$ stars in the advanced main sequence phase of evolution. In
models of such stars we typically find that the $(\ell=1\mbox{, g}_1)$ and
$(\ell=1\mbox{, p}_1)$ modes are both unstable. There are chances that
they will be detected in a number of other stars of this type. The good
luck in the case of $\nu$ Eri is that we capture it before the avoided
crossing of the $\ell=1$ modes. After it, the data on the modes yield much
less orthogonal information.

\section{Internal rotation}

\subsection{The $\ell=1$ triplets}

The frequency structure of a rotationally split $\ell=1$ mode may be
characterised by the mean splitting $S=0.5(\nu_{+1}-\nu_{-1})$ and the
asymmetry $A=\nu_{-1}+\nu_{+1}-2\nu_{0}$. If all effects of higher order
in the angular rotation rate, $\Omega$, as well as effects of a magnetic
field are negligible, then $A=0$. Both parameters, $A$ and $S$, for the
g$_1$ mode have been determined before the present campaign (Dziembowski
\& Jerzykiewicz 2003).

If we allow only an $r$-dependence in $\Omega$ then $$\nu_m=
\nu_0+{m\over2\pi}{\displaystyle\int_0^R} {dr\over R}{\cal K}
\Omega+D_0+m^2D_1.$$ The rotational splitting kernel, ${\cal
K}$, is given by

$${\cal K}=
\frac{(\xi^2_r-2\xi_r\xi_h+[\ell(\ell+1)-1]\xi^2_h)r^2\rho}
{{\displaystyle\int_0^R}{dr\over R}
[\xi^2_r+\ell(\ell+1)\xi^2_h]r^2\rho}, \eqno(1)$$ where $\xi_r$
and $\xi_h$ are defined by the following expression for the
displacement eigenvector

$$\vec\xi=(\xi_r\vec e_r + \xi_h\vec{\nabla}_h)Y_\ell^m. $$
The kernels for the two modes considered
are shown in Fig.\,6. The quantities $D_0$ and $D_1$ are quadratic in
$\Omega$. The asymmetry of the triplet arising from the quadratic effect
of rotation is $2D_1$. The quantity $D_0$ represents the frequency shift
of the $m=0$ mode. As we have mentioned in section 2.1, in principle, this
shift should be subtracted from the measured frequencies of $m=0$ modes
when we fit frequencies calculated for a spherical model.

\begin{figure}
\begin{center}
\includegraphics[width=88mm,viewport=-5 10 540 337]{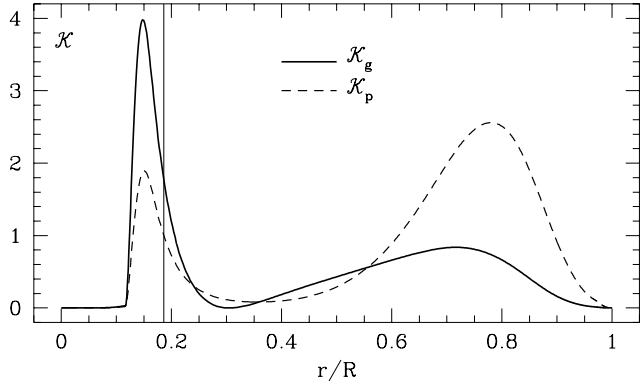}
\caption{The rotational splitting kernel for the g$_1$ and p$_1$ modes in
the model calculated with $\alpha_{\rm ov}=0$. The vertical line at
$r_{c0}/R$ marks the top of the $\mu$-gradient zone. Note the difference
between ${\cal K}$ and ${\cal E}$, shown in Fig.\,5. Rotation within the
convective core has hardly any effect on the $\ell=1$ splitting.}
\end{center}
\end{figure}

From the frequencies listed by Handler et al. (2004) for the g$_1$ mode we
derive $S_g=1.69\times10^{-2}\mbox{ cd}^{-1}$ (from here on we use
subscript $g$ to denote quantities referring to the g$_1$ mode, those
referring to the p$_1$ mode will be subscribed with $p$), which is nearly
the same as before (Dziembowski \& Jerzykiewicz 2003). There is, however,
a significant difference in the asymmetry for which the previous value was
$A_g=-7.1\times10^{-4}\mbox{ cd}^{-1}$, but from the most recent data we
now determine a value which is by more than a factor of two smaller. The
former value was larger than calculated assuming uniform rotation with the
rate inferred from $S_g$. Dziembowski \& Jerzykiewicz (2003) considered
two possible solutions. One was faster rotation in the envelope than in
the interior, the second was a contribution to $D_1$ from a hypothetical
magnetic field. The current data taken in a 158 day-long campaign and
their time distribution are insufficient for a precise determination of
$A_g$. Thus we will not use this parameter here and we will not consider
effects of a possible magnetic field. We will postpone interpretation
of the $A_g$ value until we have more precise determination from the future
data analysis.

For the p$_1$ mode we have reliable frequencies only for the $m=0$ and +1
components. The frequency of the $m=-1$ component may be estimated from
the value of $A_g$, assuming that $D_1$
arises from the second order effect of rotation with the same rate for
both modes. In this way we get
(assuming $A_g$ from earlier data which we regard as more reliable)
$$A_p=A_g{D_{1,p}\over D_{1,g}}=-1.13\times10^{-3}\mbox{ cd}^{-1}$$ and
$$\nu_{-1,p}=2\nu_{0,p}-\nu_{+1,p}+A_p=6.2250\mbox{ cd}^{-1}.$$
We stress that choosing $A_g$ from newer data or setting \mbox{$A_g =0$}
will make very small difference.
In a similar manner we find a frequency shift $D_{0,p}=7\times10^{-4}$ for
the $m=0$ mode. The number is below the adopted precision of the frequency
fit in our model construction. For the linear rotational frequency
splitting we get
$$S_p=\nu_{+1,p}-\nu_{0,p}-0.5A_p=0.01797+0.00056=0.0185{\rm cd}^{-1}$$

\subsection{Rotation rate in the envelope and in the $\mu$-gradient zone}

Since we have data on rotational splitting only for two modes, our
inference on the internal rotation must rely on simplifying
assumptions. Results from seismic sounding of the solar internal
rotation suggest that within chemically homogeneous radiative
layers the rotation rate should be close to uniform. A possible
steep gradient may occur only in the chemically inhomogeneous zone
around the convective core, where the $\mu$-gradient stabilises
differential rotation. With this in mind, we write

$$2\pi S_g=K_{c,g}\bar\Omega_c+K_{e,g}\bar\Omega_e$$ and $$2\pi
S_p=K_{c,p}\bar\Omega_c+K_{e,p}\bar\Omega_e,$$ where
$$K_{c,j}={\displaystyle\int_0^{r_{c0}}} {dr\over R}{\cal K}_j,
\qquad K_{e,j}={\displaystyle\int_{r_{c0}}^R} {dr\over R}{\cal
K}_j,$$ and $j\equiv p$ or $g$. In our fitted models we have
$$K_{c,g}=0.187 (0.168), \qquad K_{e,g}=0.345(0.395)$$
$$K_{c,p}=0.092 (0.101), \qquad K_{e,p}=0.718(0.682),$$ where
numbers in brackets refer to the model calculated with
$\alpha_{\rm ov}=0.1$. The results are nearly the same for both
models. We find $$\bar\Omega_c\approx3\bar\Omega_e.$$

Figure 6 shows the averaging kernels for $\Omega$. The fact that
the convective core virtually does not contribute to the splitting
is an exclusive property of $\ell=1$ modes which may be easily
seen by considering the behaviour of eigenfunctions near the
$r\rightarrow0$ singularity. For finite $\vec\xi$'s we have
then $\xi_r\rightarrow\ell\xi_h$, implying (see Eq.(1)) ${\cal
K}\rightarrow0$ if $\ell=1$. Thus, the value $\bar\Omega_c$
refers only to the $\mu$ gradient zone. It is clear that our
result implies a very sharp decline of the rotation in the layer
between the current core boundary and that at the ZAMS phase. How
sharp the decline is depends on the form of the $\Omega(r)$
dependence. A factor $\sim 5$ is derived if we assume a linear
decrease of $\Omega$ between $r=r_c$ and $r_{c0}$ and constant
$\Omega$ for $r>r_{c0}$.

With this simple form of $\Omega(r)$ we find an equatorial velocity of
rotation $v_e=6.1$ km/s, which is 2/3 of the value inferred from $S_g$ on
the assumption of uniform rotation. Since the quadratic effect of rotation
is still expected to arise mainly above $r=r_c$, we have a problem with
accounting for the pre-campaign value of $A_g$. The values of the
asymmetries calculated ignoring the contribution from below $r=r_{c0}$ are
$A_g=-2.0\times 10^{-4}$ and $A_p=-2.6\times 10^{-4}$. The time for a new
discussion of the triplet asymmetry will come when we have more precise
frequencies of the modes in both triplets.

\section{How the modes excited in $\nu$ Eri are driven}

Oscillation driving in B-type stars seems rather well understood (see e.g.
Pamyatnykh 1999). For low $\ell$ modes theory predicts the occurrence of
two instability domains within the main sequence band. In models of stars
with earlier spectral types (B1 - B3) low order p- and g-modes are
pulsationally unstable, as found in $\beta$ Cep stars, and in stars with
later spectral types (B4 - B8) high order g-modes are predicted unstable,
typically found in SPB stars. $\nu$ Eri is perhaps the first $\beta$ Cep
star where both kinds of modes are present (Handler et al. 2004). In the
following, we assume that the $0.432 \mbox{ cd}^{-1}$ signal in the light
curves of $\nu$ Eri is an independent pulsation mode. There is a slight
chance that it may be a combination signal, but its required order would
be much higher than those of all the observed combination frequencies of
similar amplitude. Thus we regard such an interpretation highly unlikely.

In our fitted models we find unstable modes only in the $4.0 - 6.5 \mbox{
cd}^{-1}$ frequency range. The instability measure, $\eta$, plotted in
Fig.\,7, is not a monotonic function of frequency. There is a noticeable
maximum of the instability measure near $\nu=0.5\mbox{ cd}^{-1}$ for the
$\ell=1$ sequence and at somewhat higher frequency for $\ell=2$. The
existence of this maximum suggests that pulsational instability in this
part of the eigenmode spectrum may be found upon some modification of our
models.

\begin{figure}
\begin{center}
\includegraphics[width=88mm,viewport=-5 20 560 555]{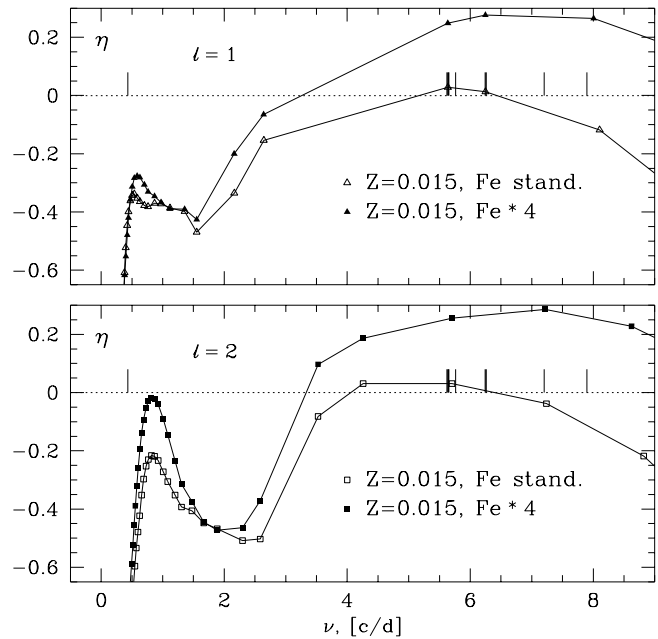}
\end{center}
\caption{The instability measure, $\eta$, for $\ell=1$ and $\ell=2$ modes
in a standard model calculated with $\alpha_{\rm ov}=0$ and in the
corresponding model with the Fe enhancement in the driving zone. $\eta >0$
for unstable modes. $\eta=-1$ if there is no active driving layer in the
star for a particular mode and it is +1 if driving occurs in the whole
interior. The position of the modes detected in $\nu$ Eri are marked with
bars on the $\eta=0$ axis.}
\end{figure}

Our proposal for extending the instability range is not original. We
follow the idea of Charpinet et al. (1996, 1997) and Fontaine et al.
(2003) that the combined effect of settling and radiative levitation may
lead to significant overabundance of iron in the driving zone. Since we do
not have a stellar evolution code treating these effects, we simply
introduced an {\it ad hoc} factor 4 enhancement of the iron-group elements
in the $\log T$ range $5.1 - 5.5$ and matched it smoothly to the standard
solar value outside. The result is shown in Fig.\,7. We may see that,
while an extension of the instability to higher order p-modes is easy,
obtaining instability in the high-order g-mode range is more difficult. We
found unstable $\ell=2$ g-modes in the model calculated with $\alpha_{\rm
ov}=0.1$, which is cooler, but only at $\nu\approx0.8\mbox{ cd}^{-1}$.
Because of the {\it ad hoc} nature of the proposed iron enhancement we
regard these results merely as an indication of a direction where the
solution of the driving problem of modes detected in $\nu$ Eri may be
found. The proposal that the diffusion effects may play a role in this
star seems not unreasonable in view of its very slow rotation which should
not induce any appreciable mixing of elements and this is the first star
where modes in such a broad frequency range were detected. However, slow
rotation cannot be the sufficient condition for a broad-band excitation.
In HD 129929, which is even a slower rotator, all modes detected so far
are confined to the narrow frequency range of 6.45 - 7 cd$^{-1}$ .

\section{Possible identification of the remaining modes}

Our manipulations on the Fe abundance in outer layers are not without
consequences for mode frequencies and hence on the seismic model of $\nu$
Eri. It is not difficult to find a model with modified Fe which fits the
three frequencies used in section 2.1. The model cannot be very different;
however, the frequencies of higher order p-modes are more significantly
changed.

Let us begin with the possible mode at $\nu=0.432\mbox{ cd}^{-1}$. Though
it is easier to find instability at higher degrees, the $\ell=1$
identification seems more plausible because in this case the maximum of
$\eta$ occurs closer to the observed frequency. Assuming $\ell=1$ and the
rotational splitting the same as in the $(\ell=1, \mbox{ g}_1)$ mode, we
obtained 0.429 and 0.435 cd$^{-1}$ for the frequency of the $(\ell=1,
m=-1,{\rm g}_{16})$ mode in the models with $\alpha_{\rm ov}=0$ and 0.1,
respectively. Thus, we regard such an identification for the
$\nu=0.432\mbox{ cd}^{-1}$ peak as likely.

For the observed variation at $\nu=7.20\mbox{ cd}^{-1}$, the only possible
identification, regardless of the choice of $\alpha_{\rm ov}$ and
modification in the Fe abundance, is a mode of the 
\mbox{$(\ell=2,\mbox{p}_{0})$}
quintuplet. In this case the effect of the Fe enhancement is more
significant. The enhancement described in the previous section leads to a
frequency decrease by 0.02, which is comparable with the rotational
splitting. With this model the best fit is for $m=1$, while with the
standard models it is for $m=2$.

The effect of the Fe enhancement increases with mode frequency. This is
not surprising because the acoustic propagation zone expands to outer
layers where the effect of the enhancement is largest. For the observed
mode at $\nu=7.90\mbox{ cd}^{-1}$ we know the $\ell$-value from
photometry, and it is 1. In our standard seismic models the $\ell=1,p_2$
modes have frequencies 8.10 and $8.03\mbox{ cd}^{-1}$ in models with
$\alpha_{\rm ov}=0$ and 0.1, respectively. In this case, even with the
freedom in choosing the $m$ value we cannot reproduce the measured
frequency. The frequency shift due to the Fe enhancement, which is about
0.1 cd$^{-1}$, brings the calculated frequencies close to the fit. We
regard this fact as partial support for the proposed effect.

We note that fitting the frequencies of all the modes detected in the 5.6
- 7.9 cd$^{-1}$ frequency range is possible in standard evolutionary
models if one allows lower effective temperature and overshooting in the
range $0.2<\alpha_{\rm ov}<0.3$ (Ausseloos et al. 2004).

\section{Conclusions and discussion}

We believe that our seismic models of $\nu$ Eridani yield a good
approximation to its internal structure and rotation but there is room for
improvement and a need for a full explanation of mode driving. We did not
fully succeed in the interpretation of the oscillation spectrum of $\nu$
Eridani with our standard evolutionary models and our linear nonadiabatic
treatment of stellar oscillations. We fail to reproduce mode excitation in
the broad frequency range, as observed, and to reproduce the frequency of
the ($\ell=1,p_2$) mode, associated with the highest frequency peak in the
spectrum. We showed that both problems may be cured by allowing an
enhancement of the iron abundance in the zone of the iron-opacity bump.
The enhancement may result from effects of radiative levitation, as first
proposed by Charpinet et al. (1996, 1997) to explain oscillations in sdB
stars. Our proposal is based on models with an {\it ad hoc} iron
enhancement in the bump zone and must be checked with use of stellar
evolution codes that take into account effects of levitation and diffusion
of chemical elements. We regard this as the most timely theoretical work
aimed at the interpretation of $\nu$ Eridani.

The proposed modification concerns only the outer layers and has a small
effect on models of the stellar interior and the frequencies of g and
p$_1$ modes. Therefore we believe that our seismic models describe the
deep internal structure of the star. These models reproduce frequencies of
the fundamental radial modes and two $\ell=1$ modes and have effective
temperatures and luminosities within the measurement error box.
Satisfactory models exist in a range of the overshooting parameter,
$\alpha_{\rm ov}=(0 - 0.12)$, which corresponds to the mass range $(9.9 -
9.0) M_\odot$, age range $(16 - 20)$ My, and a fractional hydrogen
depletion in the core range $(0.34 - 0.38)$. Here the important finding is
the upper limit for the extent of overshooting. There is a prospect for
getting a more stringent constraint on $\alpha_{\rm ov}$ with a more
accurate determination of the star's effective temperature.

We believe that our most important finding concerns internal rotation. We
presented evidence for a sharp decline of the rotation rate, $\Omega$,
through the $\mu$-gradient zone, around the shrinking convective zone.
With data on the splitting for only two $\ell=1$ modes, our estimates had
to rest on the assumed form of the $\Omega(r)$ dependence. Since a
significant gradient of $\Omega$ is possible only in the $\mu$-gradient
zone, we adopted a constant $\Omega$ above it (rotation within the
convective core has a negligible influence on the splitting). With this
simplification we found that data require a decline by a factor $\sim 5$
within the zone and an equatorial velocity of about 6 km/s. The second
order effect from this slow rotation in the envelope cannot account for
the asymmetry of the $(\ell=1\mbox{, g}_1)$ triplet as determined from the
pre-campaign data. Only with new observations we will be able to tell
whether the problem is real. Our star is not the first $\beta$ Cephei star
for which evidence for nonuniform rotation was put forward. However, we
believe that in the $\nu$ Eri case the evidence is significantly stronger
than in the case of HD 129929 (Aerts et al., 2003), because we relied on
splitting data for modes having very different probing kernels.

$\nu$ Eridani proved to be a very important pulsating star. After the Sun,
it so far is perhaps the most rewarding main sequence object for
asteroseismology. Many more modes were detected in several $\delta$ Scuti
stars but in none of them so many modes have unambiguously been
identified. This star deserves continued observational efforts. An
extended data base is needed for precise determination of the frequencies
of the modes in the $\ell=1$ triplets. This is important for a more
credible assessment of differential rotation. In this context it would be
important to obtain a precise value of the projected equatorial velocity
of rotation from spectroscopy. The best $v\sin i$ value currently available
(Aerts et al. 2004)
is higher than that of equatorial velocity inferred in this work which
may however be in part due to the combined effects of pulsational and
thermal line broadening.

\section*{ACKNOWLEDGEMENTS}

This work was supported by Polish KBN grants
5 P03D 012 20 and 5 P03D 030 20 and by the Austrian Fonds zur F\"orderung der
wissenschaftlichen Forschung under grant R12-N02.

\bsp

\end{document}